# Influence of the algorithm's reliability and transparency in the user's decision-making process

Sourabh Satish Zanwar

M.Sc. Computer Science, RWTH Aachen University

# Table of Contents

## Abstract

Algorithms have been becoming increasingly relevant for various decision-making processes in the forms of Decision Support Systems or Decision-making systems in areas such as Criminal-Justice systems, Job Application Filtering, Medicine, and Healthcare to name a few. It is crucial for these algorithms to be fair and for the users to have confidence in these decisions, especially in the above contexts, because they have a high impact on society. We conduct an online empirical study with 61 participants to find out how the change in transparency and reliability of an algorithm which determines the probability of lesions being melanoma could impact users' decision-making process, as well as the confidence in the decisions made by the algorithm. The results indicate that people show at least moderate confidence in the decisions of the algorithm even when the reliability is bad. However, they would not blindly follow the algorithm's wrong decisions.

## 1. Introduction

As a consequence of advances in the field of artificial intelligence, algorithms have become more involved in the decision-making process at various levels in many application areas, This is called Algorithmic Decision Decision Making (ADM), and the systems using ADM are called Algorithmic Decision Systems (ADS). In general, algorithms are more capable of making fair and efficient decisions, however, there is a great deal of concern regarding the confidence in their decisions and their fairness which prevents them from being used on a very wide scale. Confidence is defined by the Cambridge dictionary as the feeling of trust in someone or something (Cambridge English Dictionary, n.d.). This definition can be extended to confidence in the decision made by the algorithm as the confidence of the user in the decisions of the algorithmic decision system. We define fairness as the lack of discrimination or bias in the decisions and algorithmic transparency as openness of the data used and the algorithm's underlying model (Lepri et al, 2017). Previous research (Wang et al, 2020) found that algorithms are perceived to be fair on the basis of algorithmic outcomes, that is if the outcome is favorable or unfavorable to individual users. There has been no research on how transparency of an algorithm influences it's perceived fairness. Another factor to be considered while implementing the use of algorithms in decision making is the reliability of the algorithm, which is its quality of performing consistently well. A study (Salmons et al, 2018) shows that humans would conform less with an unreliable algorithm.

Criminal Justice systems, Employment, Insurance, Social Benefits, Medical, and Healthcare and many other areas use these Algorithmic Decision Systems. The decisions in areas like Justice systems and medical systems could be very crucial and thus could have many personal and social implications. It is necessary to find out and study the limitations of the algorithms and how the involved people perceive these decisions. According to a 2016 study, human errors in the medical scenarios are believed to be responsible for over 250,000 deaths every year in the US alone (John Hopkins Medicine, 2016). Thus using algorithms in making decisions in this context will help reduce the uncertainty and improve the accuracy, thus leading to fewer errors. There have been multiple attempts of leveraging algorithms based on Artificial Intelligence in medical systems, especially for the fields of Orthopedics (Chen et al., 2018), Cardiology (Kaur et al., 2017), and Dermatology (Esteva et al., 2017; Tschandl et al., 2020), to name a few. We focus on the application of algorithms using AI for Dermatology, diagnosis of melanoma to be specific. To identify a melanoma, an acronym ABCDEs (Qiuyu J, 2019) of melanoma was developed. Where the letters ABCDEs stand for Asymmetry, Border Irregularity, Color variation, Diameter larger than 6 mm, and Evolutions of the lesion respectively. Polesie et al. (2020) conducted a survey of  N = 1271 dermatologists and found that over 75% of them believed that using Algorithms will improve dermatologic care. In the current study, we try to find out how the reliability and transparency of the algorithm used in decision-making influence the user's confidence and perceived fairness. We also study the change in the understanding of an algorithm before and after its use.

## 1.1 Influence of transparency

According to a previous study by Rene F. Kizilcec (2016), the trust in algorithms reduced with increasing transparency after a certain point. In the current study, we try to find out how the transparency of the algorithm used in decision-making influences the user's confidence and perceived fairness. We also study if the influence of the algorithm on the user's decision changes for different levels of transparency. For these, we have the following hypothesis and research questions:

**H1.** Transparency of algorithm is positively associated with the perception of fairness

**H2.** Confidence in an algorithm is less if the users know its error rate as compared to no information or error rate

**RQ1.** How transparency of the algorithm impacts the influence of the algorithm on the user's decision making?

## 1.2 Influence of Reliability

The user's confidence in an algorithm is influenced by the reliability of the algorithm, that is how consistently the algorithm gives a correct output (Yin M. et al, 2019). We study how reliability influences perceived fairness and confidence in the decision of the algorithm. We formulate the following:

**H3.** The users show less confidence in decision made by an algorithm with low reliability

**H4.** The users have more distinct predictions from an algorithm that has low reliability.

**RQ2.** How reliability of the algorithm impacts the influence of the algorithm on the user's decision making?

## 1.3 Understanding of the algorithm

In the survey by Polesie et al. (2020), it was found that over 85% agreed that algorithms are increasingly used in their field, but only about 23% understood how the algorithm worked. Here, we also try to find if the understanding of the algorithm is influenced by its transparency and use of the algorithm in the decision-making process. We formulated the following:

**H5.** An algorithm with higher transparency is perceived to be more understandable even before using it.

**RQ3.** How does the use of algorithms in the decision-making process change its understanding?

## 2. Methodology

### 2.1 Experimental Design

We conduct an online study with 2(Transparency (Low: n = 30, High: n = 31) ✕ 2(Reliability (Bad: n = 31, Good: n = 30)) between-subject design. This helps us understand the effect of transparency and reliability on various factors that we wish to find the influence on. By varying the transparency we wish to test hypotheses H1, H2, and answer research question RQ1, whereas varying reliability for hypotheses H3, H4, and RQ2. All the participants were asked to predict the probability of presented Lesion (in %) based on their knowledge and ABCDEs of melanoma for 15 cases, out of these 5 cases were ambiguous, and the remaining 10 were split up in positive and negative for melanoma. We carried out this online survey on soscisurvey website.

### 2.2 Operationalization

#### 2.2.1 Independent Variables

**Transparency:** We have two levels of transparency, namely high and low. We operationalize this by showing the participants different data about the training of the algorithm. In the case of the algorithm with high transparency, participants are informed about many elements of the algorithm, mainly about its training, such as the algorithm was based on ABCDEs of melanomas, how the algorithm was developed, how many cases were used to train and test it, and about the error rate of the algorithm. In contrast, for the algorithm with low transparency, no information whatsoever is given.

**Reliability:** For reliability, the two levels in our study are algorithms with good reliability and bad reliability. For the more reliable algorithm, all the 15 algorithmic predictions are correct, whereas, in the case of the less reliable algorithm, 3 of the algorithmic predictions are noticeably wrong. One is False Positive while two are False Negatives.

#### 2.2.2 Dependent Variables

**The difference in Prediction, Confidence in own decision, and algorithms decision:** We ask the Participants to input the prediction(in %) of the presented lesion being a melanoma for all the displayed cases one by one. We calculate the difference between the prediction of the algorithm and the participants and take the average to get the value for this variable. For every displayed case, the participants are asked to rate their confidence in their prediction as well as the prediction made by the algorithm on a 5-point Likert scale (very unsure to very sure).

**Influence of algorithm, Attitude-to-use the algorithm and Perceived Fairness of algorithm:** After the users go through all the cases and make the predictions, we ask them to rate the following sentences 'The decisions of algorithm influenced my decision.' and 'I think it is good when decision-makers in the medical system receive assistance from algorithm-based recommendation systems.' for influence and attitude-to-use respectively on 5-point Likert scale (Strongly disagree to Strongly agree). For perceived fairness, we ask the participants to rate the algorithm for fairness on a 5-point Likert scale(very unfair to very fair).

**Evaluation of the algorithm:** To evaluate the algorithm, we ask participants 9 questions after the presentation of stimulus material. We have included the questions in the appendix section. These included sentences like 'algorithm was reliable/unreliable', 'recommendations of the algorithm were reasonable'. Four of these questions were on an inverted scale and after collection of data, we inverted the ratings. On the calculation of Cronbach's alpha to measure the internal consistency of the questions a score of 0.796 was obtained and on removing one of the questions it increased to 0.811 and thus we decided to use only those 8 questions.

### 2.2.3 Moderating and Mediating Variables

**Algorithmic appreciation:** Algorithmic appreciation, the recognition or enjoyment of the good qualities of the algorithm, is measured at the beginning of the study before presenting any stimulus material using 28 statements (of which 14 are on a reversed scaled) which are rated by participants on 7-point Likert scale (Completely disagree to Completely agree). The average of these values is used to calculate the value of this variable which is then included as a moderating variable.

**Change in understanding:** We ask the participants to rate three statements on a 7-point Likert scale before and after presenting the stimulus material. These ratings were averaged separately and used as pre and post understanding of the algorithm. We consider this to be a mediating variable.

### 2.3 Participants

For this online study, we had 61 participants. We recruited these participants primarily through mailing lists to medicine departments of various universities as well as with the help of various social media platforms. Of these 61 participants, 25 were males. Over 75% of the participants (n = 45) were either medical students or professionals who had at least basic information on ABCDEs of Melanoma, 4 of the participants had a Computer Science background. As a reward for participation, the participants could voluntarily enter a lottery of 150€(1 - 50€, 4 - 25€).

**2.4 Procedure**

When the participants opened the link for the survey they were welcomed with an explanation about the study and our motivation for the same in brief. They were also briefly informed about the structure and confidentiality of personal data. After that, they were randomly divided into 2 × 2 groups for different levels of reliability and transparency of the algorithm. Each participant is then redirected to a questionnaire for their respective case of the algorithm's reliability and transparency, where they are asked about algorithmic appreciation and then ABCDEs of Melanoma are explained to them. Here we also performed an Attention check of the participants, and then a demo case is presented. Information on the algorithm is provided to the participants. Another attention check is performed wherein the participants who are assigned the algorithm with high transparency are asked 3 questions, as opposed to 1 question for participants with the algorithm with lower transparency. This is because there is no information provided about the algorithm to the participants in case of low transparency. Then participants are asked to rate their understanding of the algorithm.

The participants are shown the 15 cases one by one and asked to predict the possibility of the displayed nevus being melanoma, as well as confidence in one's own decision and algorithm's decision. The type of cases displayed varied based on the reliability of the algorithm. After going through the 15 cases, they rated their understanding of the algorithm again, they also answered a few more questions regarding Evaluation, perceived Fairness, Influence in decision and Attitude to use the algorithm. We also asked participants to rank five fields where algorithms are used (Diagnosis of Skin Cancer, Selection of candidates in a company, Decisions in the criminal justice system, Recommendations in dating app and Triage) on basis of their severity and where algorithms should be used. A few more demographic questions were asked, and finally, we thanked and debriefed the participants, they also had a chance to take part in a lottery that was conducted as a reward for the participants.

## 3. Results

**3.1 Attitude to use the algorithm**

In general, the participants had a positive attitude towards using the algorithm-based recommendation systems for decision making in medical systems, the mean score was 3.95 out of 5. On performing Two-Way ANOVA analysis, there was no significant relationship observed between both the transparency of the algorithm with the attitude to use the algorithm (Low Transparency: M = 3.87, SD = .860; High Transparency: M = 4.03, SD = .752) and the algorithm's reliability with the attitude to use the algorithm (Bad Reliability: M =

3.83, SD = .869; Good Reliability: M = 4.06, SD = .742), this can be seen in Table 1 below.

**Table 1**
Attitude to use the algorithm

| Source | df | F(1.57) | Sig. | Partial Eta Square |
|---|---|---|---|---|
| Transparency | 1 | .658 | .421 | .011 |
| Reliability | 1 | 1.235 | .271 | .021 |
| Transparency * Reliability | 1 | .029 | .865 | .001 |

## 3.2 Perceived Fairness of Algorithm

The algorithm was, in general, perceived to be fair by the participants(M = 3.62, SD = .697), irrespective of the variations in levels of transparency and reliability. We could not find any significance in the impact of transparency of the algorithm on the perception of its fairness, this can be seen in Table 2, which shows results of Two-Way ANOVA analysis. This led to the rejection of the hypothesis H1.

**Table 2**
Effect of transparency and reliability on Perceived fairness of the algorithm

| Source | df | F(1.57) | Sig. | Partial Eta Square |
|---|---|---|---|---|
| Transparency | 1 | 1.036 | .313 | .015 |
| Reliability | 1 | .011 | .917 | .038 |
| Transparency * Reliability | 1 | 1.036 | .313 | .001 |

## 3.3 Influence of the Algorithm

According to the response of the participants, we performed univariate ANOVA and found that there was no significant difference between the influence of the algorithm on the participant's decision for different levels of transparency and reliability. To answer our research questions RQ1 and RQ2, we found **no influence** of the algorithm's transparency and reliability on the user's decision-making., this can be seen in Table 3 below. In general, there was an above-average influence of the algorithm's decision on the participant's decision ( M = 3.26, SD = .890).

**Table 3**
Effect of transparency and reliability on the Influence of the algorithm

| Source | df | F(1.57) | Sig. | Partial Eta Square |
|---|---|---|---|---|
| Transparency | 1 | 1.464 | .231 | .025 |
| Reliability | 1 | .302 | .584 | .005 |
| Transparency * Reliability | 1 | .711 | .403 | .012 |

### 3.4 Deviation in Prediction

We found that the prediction of participants for the melanoma is different from the algorithm's prediction by an average of about 17% irrespective of the different variations of the algorithm. To see if some significant effect exists we performed Two-Way ANOVA on absolute deviation in means of predictions between algorithm and users, the results of which can be seen in Table 4. There is a significant effect of the Reliability of the algorithm on the deviation in prediction (p-value = .001). There was an increase in the deviation by about 6% between algorithms with bad reliability(M = 20.11, SD = 6.16) and good reliability(M = 14.24, SD = 7.35). Based on these results we accept hypothesis H4, as there is more deviation in participant's decisions from the algorithm's decision in case of a less reliable algorithm.

**Table 4**
Effect of transparency and reliability on the deviation between predictions made by algorithm and predictions made by participants

| Source | df | F(1.57) | Sig. | Partial Eta Square |
|---|---|---|---|---|
| Transparency | 1 | .414 | .522 | .007 |
| Reliability | 1 | 11.168 | .001 | .164 |
| Transparency * Reliability | 1 | 1.310 | .257 | .022 |

### 3.5 Evaluation of the Algorithm

The participants evaluated the algorithm to be just above average, the mean rating for the algorithm's evaluation is found to be 4.66 out of 7. Also, there is no significant relationship found on performing Two-way ANOVA on evaluation by transparency and reliability of the algorithm. The participants evaluated the algorithms of different conditions similarly, this can be observed in Table 5.

**Table 5**
Evaluation of the algorithm

| Transparency | Reliability | Mean | Std. Deviation | N |
|---|---|---|---|---|
| Low Transparency | Bad Reliability | 4.733 | .873 | 15 |
|  | High Reliability | 4.741 | .688 | 15 |
|  | Total | 4.737 | .772 | 30 |
| High Transparency | Bad Reliability | 4.453 | .845 | 16 |
|  | High Reliability | 4.741 | .707 | 15 |
|  | Total | 4.529 | .782 | 31 |
| Total | Bad Reliability | 4.588 | .865 | 31 |
|  | High Reliability | 4.741 | .686 | 30 |
|  | Total | 4.663 | .774 | 61 |

## 3.6 Confidence in the algorithm's and one's own decision

We found that participants showed high confidence in both their own decision as well as the algorithm's decision. With mean confidence in their own decision at 3.57 and mean confidence in the algorithm's decision at 3.48, there is not much difference in confidence. There is also no significant relationship between the transparency and reliability of the algorithm with confidence. We can see the statistical data for Two-way ANOVA analysis on both of these in the following Table 6. Taking into considerations we have to reject hypothesis H2 as the confidence in decisions of the algorithm was high even when the error rates of the algorithm were known in case of high transparency(M = 3.43, SD = .398). We also reject hypothesis H3 because high confidence is shown even when the algorithm is less reliable(M = 3.38, SD = .443).

**Table 6**
Effect of transparency and reliability on confidence in one's own decision and algorithms decision

| Source | Algorithm's Decision | | | | Self-decision | | | |
|---|---|---|---|---|---|---|---|---|
|  | df | F(1.57) | Sig. | Partial Eta Square | df | F(1.57) | Sig. | Partial Eta Square |
| Transparency | 1 | .839 | .363 | .015 | 1 | 1.626 | .207 | .028 |
| Reliability | 1 | 2.243 | .140 | .038 | 1 | .573 | .452 | .020 |
| Transparency * Reliability | 1 | .063 | .802 | .001 | 1 | .958 | .332 | .017 |

### 3.7 Understanding of the Algorithm

The mean understanding of the algorithm before the trials was found to be 4.58/7.00, SD = 1.28. On performing Two-way ANOVA with repeated measures on the pre and post -understanding of the algorithm for different levels of transparency we found a significant effect with significance level of $p<.1$, this can be seen in Table 7. Based on this we can accept the hypothesis H5 which states that the understanding of the algorithm with high transparency is higher as compared to low transparency, we can also see this in Figure 1. We did not find a large difference in the understanding of the algorithm before(M = 4.58, SD = 1.28) and after(M = 4.88, SD = 1.31) the trials, this answers RQ3.

**Table 7**
Effect of transparency and reliability on understanding of algorithm (pre-trial)

| Source | Significance | Partial Eta Square |
|---|---|---|
| Transparency | **.045** | .068 |
| Reliability | .912 | .00 |
| Transparency * Reliability | .340 | .016 |

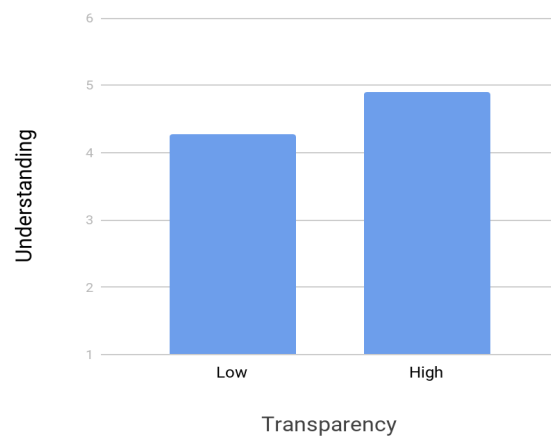

**Figure 1.** Mean understanding of algorithm before trials

## 4. Discussion

With this study, we explore the use of algorithms for decision making for the detection of melanoma. From the results of the online survey, we find that generally, people have a positive attitude towards using the algorithms in the decision-making process. Not only a positive attitude but the participants recorded high confidence in the decision of the algorithm. Even when the algorithm performed with bad reliability, at least moderate confidence was recorded, however, people do not blindly follow the decision of the algorithm. In contrast to the high confidence in a less reliable algorithm, we observed more deviation in the decisions made by the participants and the algorithm for the algorithm with bad reliability, we can observe this in Figure 2. This is similar to the results obtained by Salmons et al (2018). From the data we collected, only 10 of the 61 participants were against using the algorithmic recommendations, from this we can say that people, in general, are in favor of using them, this is in line with the survey conducted by Polesie et al (2020). However, from reading the reasons for their answers, we find that a lot of them have concerns about the data on which the algorithms were trained, as well as the accuracy of the algorithms. Although we include perceived Fairness, it can be argued that it might not be applicable in the case of the detection of melanoma.

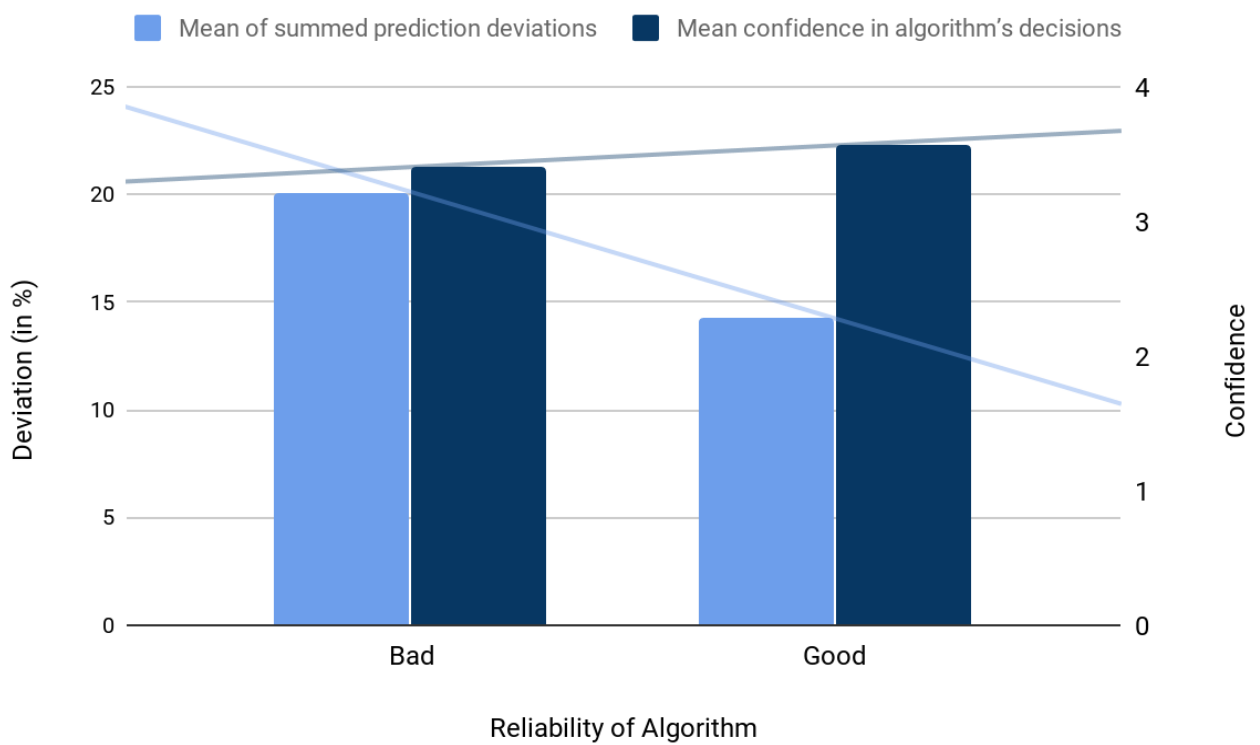

**Figure 2.** Effect of reliability of the algorithm on Deviation in predictions made by algorithm and participant and confidence in the algorithm's decision

## 4.1 Limitations

The most important limitation of our study is the number of participants. There could be different results if these studies are carried out with more participants. Another limitation could be that our manipulations for the levels of transparency and reliability are not strong enough. As there were a lot of cases that participants had to rate, it could be possible that the participants would forget the error rate in case of a more transparent algorithm. Whereas in the case of reliability, the ratio of bad cases was very less as compared to good cases, more bad cases might cause a change in the recorded confidence of the confidence with bad reliability by participants. All through the survey, the participants were presented with a lot of information, this could be overwhelming for a few participants, in this case, a lab study with a longer duration will give better results as the information would not be overwhelming over a longer period.

## 4.2 Future Work

Apart from the data that we used in this paper, we also recorded a lot of data that could be very useful in finding significant relations along with considering the confounding variables. As stated in the limitations, our manipulations in reliability and transparency might not be as good, and a study with better manipulations could be useful in finding better results. Along with better manipulations, the levels of manipulations can also be increased. Similar studies can be carried out for various other contexts where algorithmic recommendations can be used for decision making such as criminal justice systems, recommendations in dating apps, to name a few.

Appendix

1. PostEvaluation of Algorithms from the survey

I have largely ignored the algorithm in my decisions

The algorithm was very helpful in the decision making process.

I have incorporated the recommendations of the algorithm into my decision-making process.

I found the recommendations of the algorithm reasonable.

The recommendations of the algorithm were in line with my assessment.

The recommendations of the algorithm were easy to understand.

In my opinion the algorithm did not give good recommendations.

The algorithm made errors.

The algorithm was unreliable.